# Study on Intelligent Forecasting of Credit Bond Default Risk

KAI REN[1]

## ABSTRACT

Credit risk in China’s bond market has become increasingly evident, creating an escalating risk of default for investors. Given the current incomplete and inaccurate bond information disclosure, timely tracking and forecasting the individual credit bond default risks have become essential to maintain market stability and ensure healthy development. This paper proposes an Intelligent Forecasting Framework for Default Risk that provides precise day-by-day default risk prediction. In this framework, we first summarize the factors that impact credit bond defaults and construct a risk index system. Then, we employ a combined default probability annotation method based on the evolutionary characteristics of bond default risk. The method considers the weighted average of Variational Bayesian Gaussian Mixture estimation, Market Index estimation, and Default Trend Backward estimation for daily default risk annotation of matured or defaulted bonds according to the risk index system. Moreover, to mine time-series correlation and cross-sectional index correlation features efficiently, an intelligent prediction model for Chinese credit bond default risk is designed using the ConvLSTM neural network and trained with structured feature data. The experiments demonstrate that the predicted individual bond risk is slightly higher and substantially more responsive to fluctuations than the risk indicated by authoritative ratings, thereby improving on the inadequacies of inflated and untimely bond ratings. Consequently, our study's findings offer multiple insights for regulators, issuers, and investors.

The introduction of unsecured short-term financing bond in 2005, which relied entirely on the issuer's own credit, marked the beginning of the genuine China’s credit bond market. As of December 31, 2022, the stock size of credit bonds reached RMB 58.40 trillion, accounting for 16.97% of the stock of social financing of RMB 344.21 trillion[2], providing significant financial support for the development of the Chinese economy and occupying a vital position in the country's financial market. However, during the market's rapid expansion, credit risks have accumulated, with significant potential for hidden defaults. The material default of the "11 Chaori Bond" in 2014 marked the end of the persistent "rigid payment" situation in the credit bond market, with defaults occurring repeatedly thereafter.

---

[1] Instructor: **Xuebin Chen**. Chair Professor, School of Economics, Sichuan University.

[2] All data used in the study are sourced from the **iFind** platform.

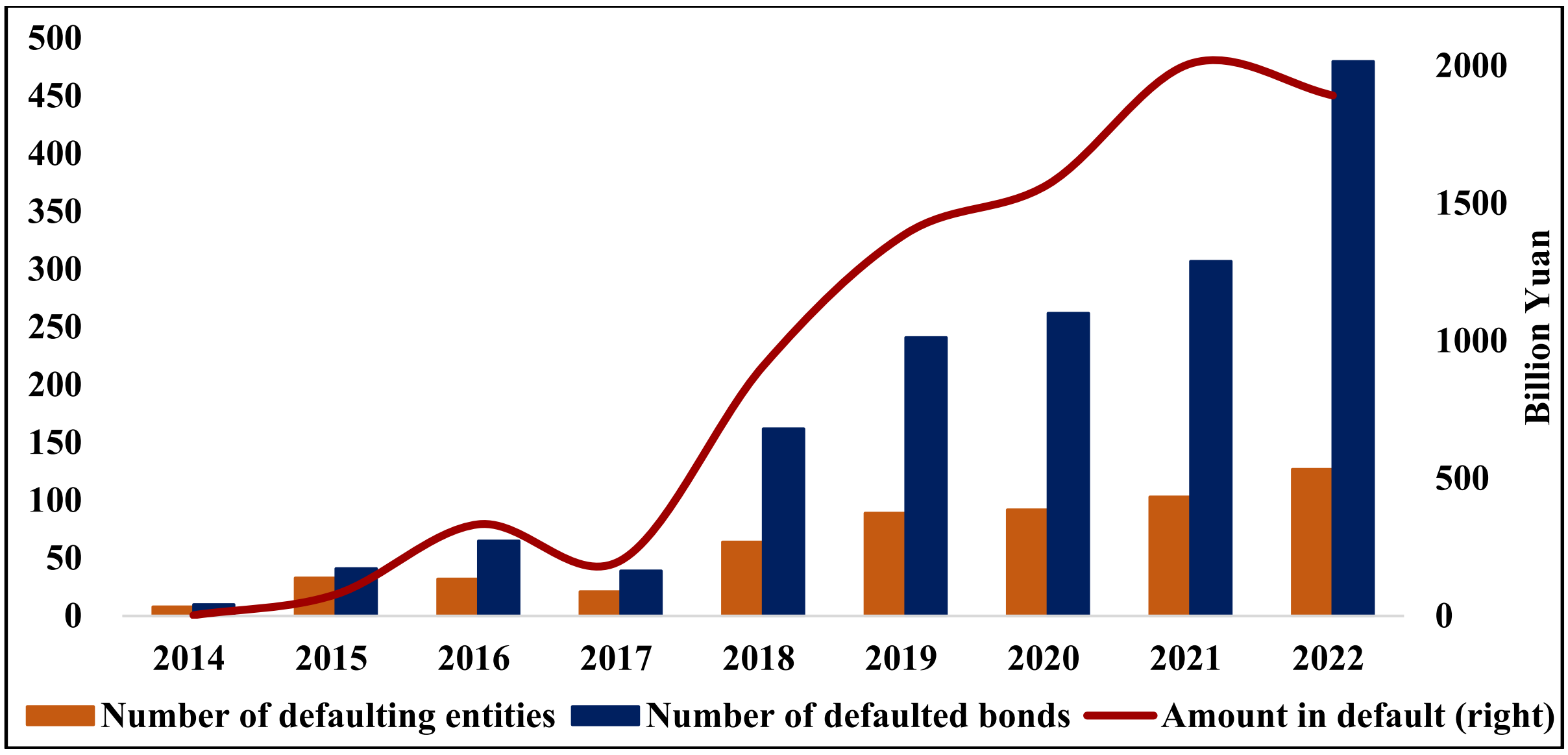

**Figure 1. Basic information on defaults in China's credit bond market, 2014-2022.**

As shown in Figure 1, the number of defaulted bonds has gradually increased since 2014, with amount in default rising rapidly and credit risk being increasingly exposed. In 2022, the number of defaulted bonds reached 480, involving 127 entities and nearly RMB 200 billion in default amount, marking a record high cumulative default rate and further normalizing credit bond defaults. However, the current market lacks a timely and effective default warning mechanism, and credit ratings, which are supposed to be a useful tool for dynamic measurement of credit bond default risk, have in some cases misled investors' judgement. Dozens of the credit bonds that defaulted in 2022 were issued with a AAA bond rating that remained at the highest rating level prior to the default, such as the bonds of Rongchuang Real Estate Group and the medium-term notes of Shanghai Shimao. It was these "safe" bonds, which were rated highly, that suddenly defaulted on a large scale, causing huge losses to investors as well as having a major impact on the credit bond market.

The surge in default risk can be primarily attributed to the pending improvement of the bond market infrastructure and the serious information asymmetry between bond issuers and investors (Liang, 2022). (1) Many bond issuers are non-listed companies, which lack complete and standardized financial information disclosure. Of the 1,607 defaulted bonds between 2014 and 2022, only 786 bonds have relatively complete financial reports. Listed bond issuers likewise sometimes distort their financial statements; (2) The ratings of bonds are often inflated and lagged in adjustment, making it difficult to accurately and timely measure credit risk. Since lower-rated bonds cannot be issued in the strict bond market, issuers often strive to obtain high ratings. For the credit rating of the same bond, the results given by domestic rating agencies are significantly higher compared to the rating results of the three authoritative international rating agencies. As shown in Figure 2, 42.32% of the credit bonds defaulted during this period had no issue rating, while those with issue ratings are all AA-rated or above, with AA+ rating being the most common, and AAA-rated bonds accounted for the largest proportion of the defaulted amount. At the same time, a large number of bonds lacked timely adjustment of tracking ratings, and sudden defaults on highly rated bonds occurred now and then.

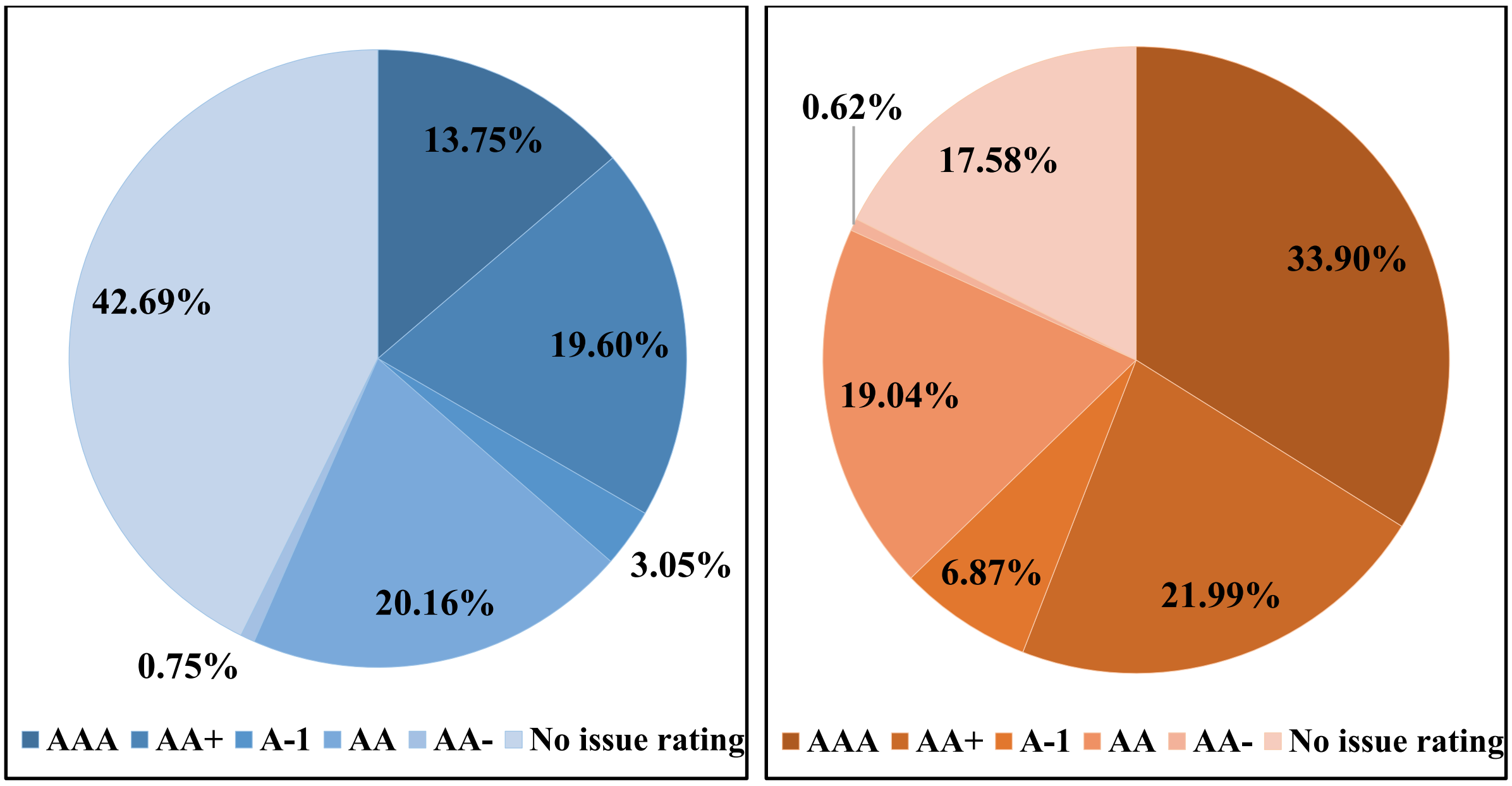

(a) (b)

**Figure 2. Distribution of all defaulted credit bond issue ratings from 2014 to 2022.** Pie chart (a) shows the percentage of the number of defaulted bonds at each issue rating. Pie chart (b) shows the percentage of default amount at each issue rating.

Therefore, in the rapidly developing credit bond market, it is important for investors to deeply explore the limited disclosure information of issuers of the bonds and a large amount of other relevant data(Foster et al., 1998). This will allow for the establishment of a timely and effective bond default risk analysis and a prediction mechanism, which can provide advanced warning of potential defaults, contributing to the healthy development of China's credit bond market.

Regarding current relevant research, classical default risk quantification models like by Reduced-Form Model and KMV, as well as deep neural network models, have proven to be more accurate at predicting defaults for individual listed companies with complete and effective information disclosure. However, the China's credit bond market studied in this paper contains a large number of bonds issued by non-listed companies, and with limited high-quality public data, the similar methods tend to be distorted (Chen et al., 2021). Moreover, the features used in previous studies are mostly focused on company financial information, which is limited by the frequency of feature data, and the credit risk prediction using the model is often limited to quarterly or lower frequency default prediction. In addition, due to the lack of day-by-day default probability labels, the supervised machine learning model can often only produce a binary judgment of whether to default but cannot give an exact estimate of default probability(Wang et al., 2019), which cannot achieve the requirement of timely and effective reflection of default risk. Furthermore, traditional machine learning models have difficulty fully extracting the combined information of both temporal and cross-sectional index association dimensions of credit bond data. For example, the Boosting model can fully extract the association information of feature index dimension but is weak in processing the temporal dimension, while the temporal characteristics of financial data are very important, the Long Short Term Memory (LSTM) model is just the opposite. This can lead to constrained prediction accuracy (Zhao et al., 2019).

To address the above problems, this paper presents an Intelligent Forecasting Framework for Default Risk covering index designing, labeling and model building for the dynamic data series of China's credit bond market. Firstly, we collect and analyze the public data of major defaulted credit bonds since 2014, and design a risk index system containing 53 specific features based on seven dimensions, such as the basic financial status of the issuer, the industry and region it belongs to, and the macroeconomy in the same period. The frequency of each feature is unified to the daily frequency to characterize the information of individual bonds as comprehensively as possible. Secondly, according to the features of credit bond default risk evolution, the Variational Bayesian Gaussian Mixture estimation, Market Index estimation and Default Trend Backward estimation are used to carve out the daily frequency default probabilities of matured and defaulted bonds as reasonable labels for deep neural network training, validation and testing. Finally, we build and train an intelligent prediction model for default risk based on the ConvLSTM deep neural network, which is capable of processing information in both time and space dimensions, to achieve the accurate day-by-day risk warning. Our framework represents a significant addition to prevailing study on deep learning in finance and possesses essential practical implications for early identification and prevention of credit bond default risk.

The rest of the paper is organized as follows. Section I reviews the related literature. Section II outlines the methodology used for data collection and risk index system construction. Section III annotates the bonds with daily frequency default probabilities. Section IV constructs the default probability prediction model using ConvLSTM and completes the training. Section V evaluates and compares the prediction results of the model. Section VI concludes the study and includes analysis for insights gained.

## I. Related Literature

In the mid-to-late 20th century, credit risk measures underwent quantitative analysis, leading to the emergence of several **classical default prediction models** in academia and industry. Based on the univariate early warning model proposed by Beaver (1966), Altman (1968) developed a 5-variable Z-Score model to predict corporate bankruptcy risk. Merton (1974) set the base for a structural model that measured credit risk using the market value of firm assets as the underlying and liabilities as the strike price, treating firm value as a European call option. Jarrow and Turnbull (1995) proposed a Reduced-Form Model by introducing "default intensity" to address the limitations of structural models in real-world applications. The LKR signal analysis model proposed by Kaminsky et al. (1998) provided an important reference for the industry's risk warning due to its high simplicity and accuracy. In the beginning of the 21$^{st}$ century, extended models with greater applicability, such as through credit derivatives market prices, were developed and widely used. Domestic studies on default risk measures initially focused on the empirical evidence and improvement of the above models. Chen (1999) verified the effectiveness of scoring models in forecasting financial deterioration of Chinese listed companies. Cheng et al. (2000) and Wu et al. (2001) utilized the KMV model in the analysis of loan default risk for Chinese listed companies. Recently, bond default risk has gained extensive attention, Yao et al. (2018) assessed the impact of traditional financial index, firm characteristics and local environmental indicators on credit bond default using a discrete-time risk model, Zhang (2019) utilized a logistic model to predict whether a firm defaults. Although these methods verified some factors' degree of influence on bond defaults, the prediction accuracy is not high.

With the continuous development of artificial intelligence technologies, **deep learning** has demonstrated superior performance in high-dimensional data processing, and extraction temporal and spatial feature, thus gaining widespread attention in the field of finance. Studies on deep learning in risk management have mostly centered on early warning systems for credit risk (Ozbayoglu et al., 2020). Chatzis et al. (2018) reported that deep neural networks can significantly improve the accuracy of risk prediction when examining risk propagation channels and early warning mechanisms in capital markets. Golbayani (2020) comprehensively compared the performance of multilayer perceptrons (MLPs) with traditional models in rating prediction tasks, revealing that MLPs provide high flexibility and accuracy. Traditional machine learning models are also applied to predict the binary judgments about the future default status of a company or borrower (Wang et al., 2019; Wang et al., 2022). In contrast, LSTM can learn long-range temporal dependencies by introducing gating mechanisms, and is widely utilized to financial asset price and earnings forecasting (Rather, 2015; Minami, 2018;). Based on the LSTM, Zhao et al. (2019) constructed a hierarchical time series fusion prediction model to forecast daily trading volume and the number of trades. Concerning risk prediction, Chen et al. (2021) employed LSTM to examine credit bond default risk and found that the model offered more accurate predictions of future default judgements. However, these models often have difficulty fully leveraging their performance in day-by-day default probability prediction due to the lack of high-quality structural data and the absence of daily frequency default probability labels.

## II. Data Collection and Risk Index System Construction

### *A. Samples and Data*

Chinese bonds can be categorized into three types based on their risk profile: interest rate bonds, quasi-interest rate bonds, and credit bonds. Credit bonds, which are high risk instruments due to their nature of being unsecured and credit-based, can be further classified into financial bonds and non-financial bonds based on the issuing entities. Financial bonds are usually issued by financial institutions such as commercial banks and securities companies with lower probabilities of default, while non-financial bonds consist of diverse instruments like corporate bonds, enterprise bonds, private placement notes (PPNs) with a higher probability of default. Only five of all defaulted credit bonds from 2014 to 2022 are financial bonds, accounting for less than 0.5% of the total, and the rest are non-financial bonds. The reasons behind such non-financial defaults are mainly related to corporate refinancing effectively affected by liquidity stratification (Zhong et al., 2021). As shown in Figure 3, the defaulted non-financial bonds are mainly corporate bonds, medium-term notes (MTN), asset-backed securities (ABS), and commercial paper. We also focus on such credit bonds to ensure comprehensiveness while reducing the complexity of data collection.

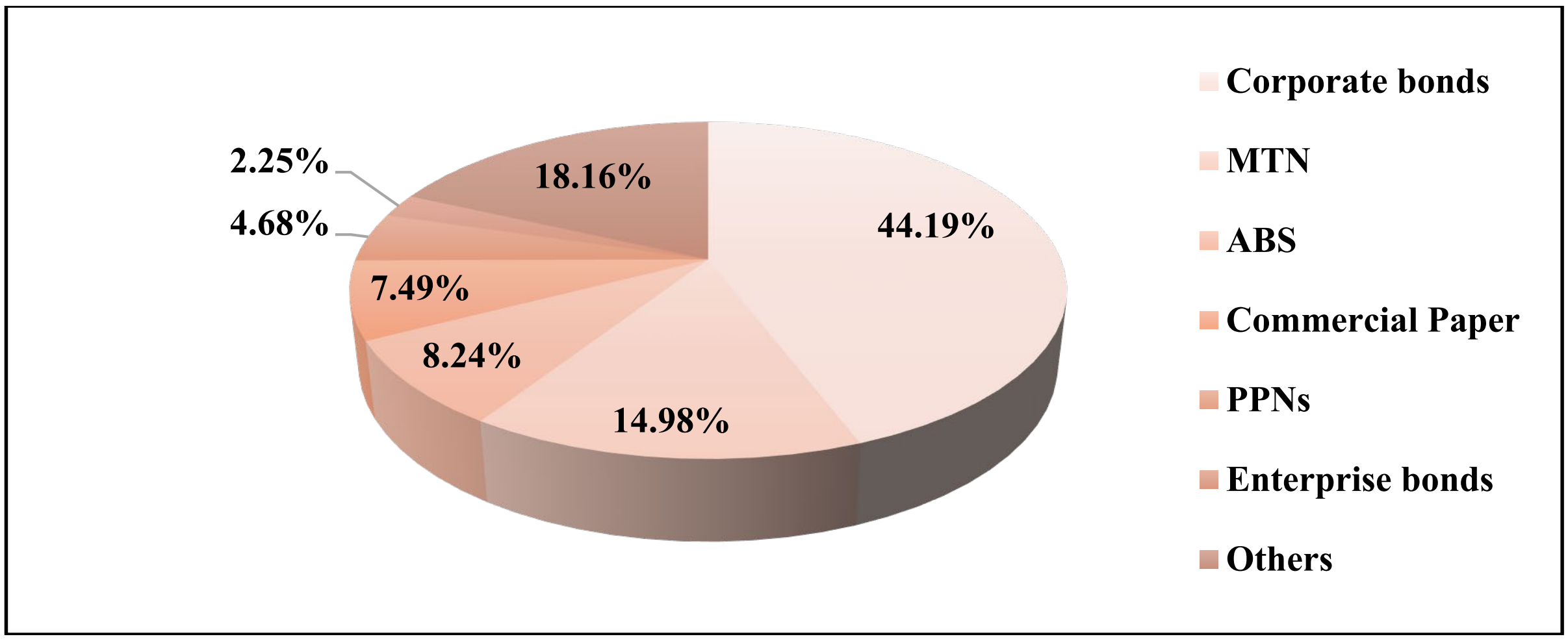

**Figure 3. Percentage of non-financial bonds in default by type, 2014 to 2022.**

Considering the data availability, we collect all matured or defaulted credit bonds issued from January 2014 to December 2022 in China's credit bond market through the iFind platform by excluding irrelevant data. This yields a total of 10,476 matured bonds and 763 defaulted bonds. Since matured bonds provide a low credit risk sample as they are guaranteed not to default, they are chosen as such. To balance high credit risk and low credit risk samples, and given the limited number of defaulted samples, we include 312 bonds that are not yet mature but rated below A to increase the number of samples with higher credit risk. We also exclude bonds with missing information, such as financial and transaction data of bond issuers. Therefore, the final sample comprises 7,361 credit bonds, including 675 high credit risk bonds (548 defaulted bonds and 127 bonds rated below A) and 6,686 low credit risk bonds that have matured, with a total of about 4.69 million daily transaction frequency data.

## *B. Credit Bond Risk Index System*

This paper comprehensively characterizes the credit features of individual bonds by summarizing the factors that affect bond defaults based on recent studies by Zhou (2020) and Xu (2022). For one thing, external macroeconomic and industry fluctuations are among the factors that contribute to credit defaults. For another, deteriorated bond repayment ability caused by internal operational mistakes, the decline in profitability and abnormal basic financial indicators led by governance deficiencies also contribute. Based on the screening criteria provided by Golbayani (2020), and considering the availability of relevant indexes, we ultimately select 53 specific features from seven dimensions: Macroeconomy, Industry and Region, Basic Financials of Issuing Entities, Bond Repayment Ability, Profitability, Other Issuer Characteristics, and Bond Market Conditions. These features are used to construct the credit bond risk index system shown in Table I.

Most of the indexes used in this study are typical macroeconomic or company fundamentals, which will not be further elaborated upon. In order to measure the level of industry and regional prosperity, this paper considers "Default probability by category" and "Default probability by region". These respectively refer to the accumulated default probability of credit bonds within the industry or region to which the bonds belong on a given trading day. Given that the shortest interval between corporate financial reports is quarterly, which has a certain lag, the index system is supplemented by collecting forecasts of major financial indicators issued by some companies, introducing "Forecast profit change" into the index system. In addition to formal financial reporting indexes, the credit-granting and guarantee situations of enterprises also play a certain role in forecasting their default status. Therefore, three indexes including "Credit residual ratio" are chosen. We also introduce "Prior default probability" to represent credit risk variability, which will be calculated later.

**Table I**

**Credit Bond Risk Index System**

| Dimension | No. Index Name | Dimension | No. Index Name |
|---|---|---|---|
| Macroeconomy | 1. Leading economic index | Bond Repayment Ability | 27. Current ratio |
| | 2. Manufacturing PMI | | 28. Quick ratio |
| | 3. PPI month-on-month (mom) | | 29. Superquick ratio |
| | 4. CPI mom | | 30. Assets-liabilities ratio |
| | 5. GDP quarter-on-quarter | | 31. Equity ratio |
| | 6. RMB to USD exchange rate | | 32. Bond to tangible assets ratio |
| | 7. 3-month Shibor Rate | Profitability | 33. Gross sales margin |
| | 8. Treasury rates for the same period | | 34. Net profit margin on sales |
| | 9. Stock of social financing scale | | 35. Return on assets |
| Industry and Region | 10. Subordinate ShenWan primary | | 36. Operating profit margin |
| | 11. Default probability by category | | 37. Return on equity |
| | 12. Default probability by region | | 38. Operating cycle |
| Basic Financials of Issuing Entities | 13. Operating revenue | | 39. Inventory turnover ratio |
| | 14. Operating cost | | 40. Receivables turnover ratio |
| | 15. Total profit | | 41. Current asset turnover ratio |
| | 16. Current assets | | 42. Equity Turnover |
| | 17. Non-current assets | | 43. Total Asset turnover |
| | 18. Total assets | Other Issuer Characteristics | 44. Forecast profit change |
| | 19. Current liabilities | | 45. Credit residual ratio |
| | 20. Non-current liabilities | | 46. Change in credit mom |
| | 21. Total liabilities | | 46. Guaranteed credit ratio |
| | 22. Total stockholders' equity | | 47. Stock price fluctuations |
| | 23. Cash flow from operations | Bond Market Conditions | 48. Trading volume |
| | 24. Cash flow from investment activities | | 49. Residual maturity |
| | 25. Cash flow from financing activities | | 50. Yield to maturity |
| | 26. Total cash flow | | 51. Risk spread |
| | | | 52. Prior default probability |

This index system has strong explanatory power in most cases of bond default. For instance, when examining the Yongcheng commercial paper default in November 2020, an analysis of changes in such variables as "Return on assets", "Total liabilities", and "Cash flow from financing activities" can indicate significant operational difficulties experienced by the company within the current quarter. This is observed during the gradual transmission of macroeconomic slowdown caused by the impact of the epidemic, which led to declining revenue in addition to industry risks caused by the weak demand for energy, and which began to further emerge in the second half of 2020. Consequently, prior to the default, market variables such as the "Risk spread" and "Yield to maturity" play an increasing role in risk disclosure. The entire process exemplifies the successful identification of risk evolution trajectory by our index system.

## III. Daily Frequency Integrated Default Probability Annotation

Currently, deep learning models used in financial data series analysis are predominantly supervised learning models (Golbayani, 2020). In other words, these models require supervision from accurate labels to calculate the loss function, back-propagate, and update parameters. However, there exists no individual bond default probability data available in China's credit bond market at present. In this section, we will detail the combined use of Variational Bayesian Gaussian Mixture estimation, Market Index estimation, and Default Trend Backward estimation to label the historical daily frequency default probabilities of collected credit bonds. With such approach, a Chinese credit bond dataset with daily frequency default probability labels can be constructed.

### *A. Variational Bayesian Gaussian Mixture estimation*

Liang (2022) and Xu (2022) investigated how external factors, financial characteristics of the firm, and bond attributes affect default risk. They found that bonds tend to exhibit some convergence in these indexes, implying that unsupervised clustering can classify bonds to estimate default probabilities. K-means, initially proposed by MacQueen (1967), is a straightforward and effective clustering method. The Gaussian Mixture Model (GMM) extends K-means and offers greater stability in inference. Additionally, its correlation assumptions align more closely with financial data (Androniceanu et al., 2020).

The GMM assumes that all the samples originate from a finite number of Gaussian distributions with unknown parameters. The central information of each obscured model is solved by using a Variational Bayesian interface. If the intended number of clusters is $K$, an optimization process of the GMM contains $K$ Gaussian distributions and the parameter $\theta=\{p,\ \mu,\ \Sigma\}$, is necessary to achieve the maximum value of the objective function $Q(\theta)$ when constrained by $\Sigma_{k=1}^{K} p_k = 1$. $p$ represents the probability distribution of the dependent variable; $\mu$, $\Sigma$ indicate the mean and covariance matrix. Specifically:

$$\theta^* = arg\max_{\theta}\ Q(\theta) \text{ and } Q(\theta) = \Sigma_{i=1}^{N} \ln\left(\Sigma_{k=1}^{K} p_k N(x_i|u_k, \Sigma_k)\right) \tag{1}$$

To align with the bond ratings, we set the target number of clusters as 22. The default risk of a bond gradually decreases as the bond rating characterized by the clustering result increases from 1 to 22. Clustering is performed using a Variable Bayesian Gaussian Mixture model[3] on 52 features, except for "Prior default probability," with 200 iterations as the maximum number of iterations. The results reveal daily frequency grade 17 (corresponding to a credit rating of AA-) and 21 (AAA) have the highest proportion among clusters, at 38.13% and 13.66%, respectively. Grade 1, representing credit rating D, has a relatively high proportion of 1.66%. This distribution is a more reasonable reflection of the actual credit level of China's credit bonds: on the one hand, the overall default risk is low, with 94.67% of A-rated and above, and less than 0.1% of the higher risk B-rated; on the other hand, once the bond-related indexes are in question, bonds will quickly slide to the high default probability of D-rated.

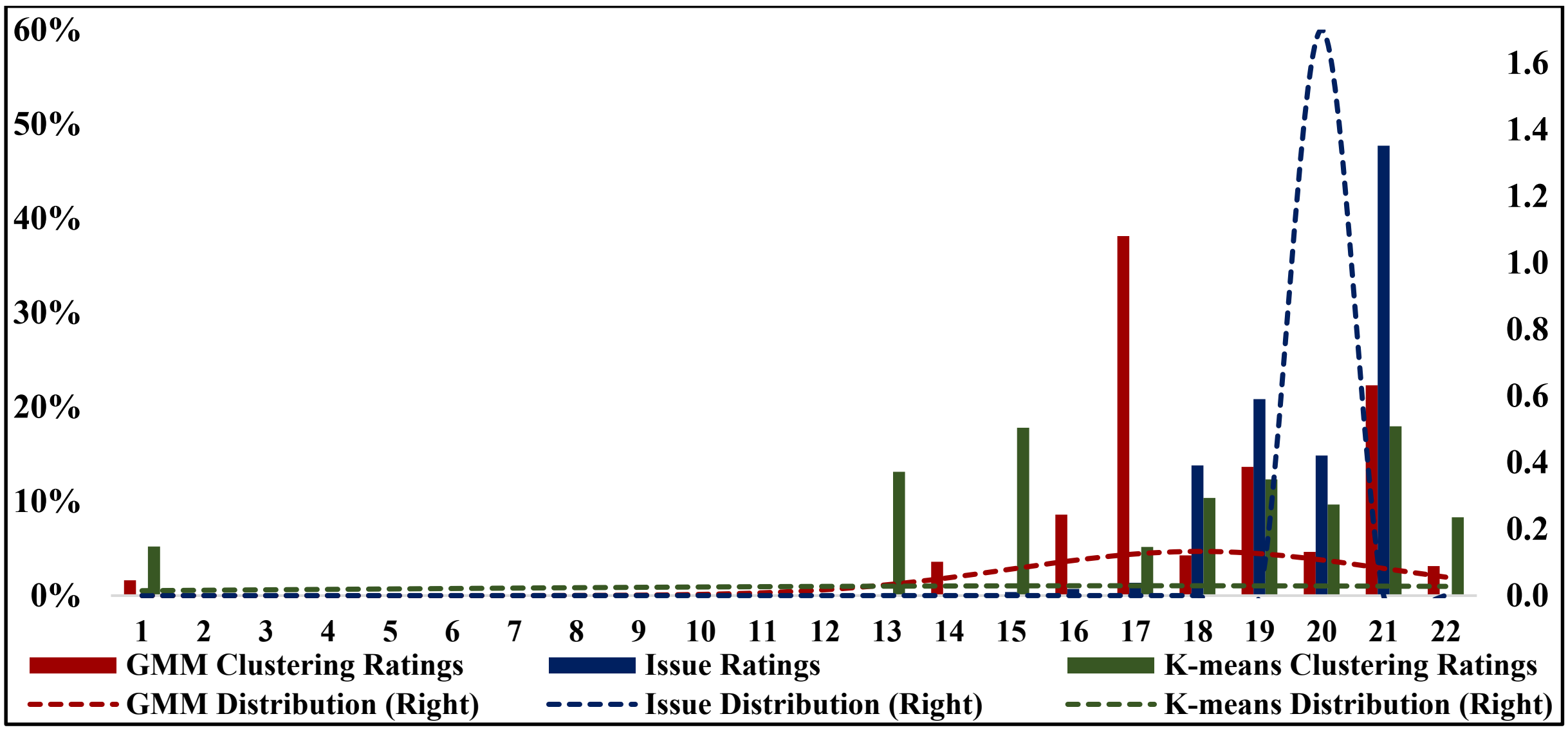

**Figure 4. Comparison of the proportion of clustering ratings and issue ratings of Chinese credit bond with the fitted normal distribution.**

Figure 4 represents the issue rating and the estimated ratings obtained from two different methods: Variational Bayesian Gaussian Mixture model and K-Means model. A normal distribution (dashed line in the Figure 4) is fit on the original grading results to facilitate a clear visual comparison of the distribution differences among three rating resuls. It can be seen that the issue ratings are concentrated around AAA- grade, whereas the distribution of the estimated ratings based on our model, with a mean value around AA, shows higher variance. This not only corrects the deviation of high issue ratings but also accurately represents the gradual increase of bond credit risk in China's credit bond market from issuance to maturity. In contrast, the K-Means model exhibits an near-uniform distribution with high variance. Such outcome diverges from the actual distribution, demonstrating the efficacy of our selected methodology.

On this basis, to establish a more precise and objective relationship between default category and default probability, the ***logistic*** function is employed to map clustering discrete variables to default probability continuous variables:

$$\hat{p}_{it}^{gmm} = \frac{1}{1 - e^{-g\left(\hat{k}_{it}^{gmm}\right)}}, \tag{2}$$

[3] This is implemented using the ***mixture.BayesianGaussianMixture*** module in the ***Keras*** toolkit.

where $i$ denotes individual bonds, $t$ represents trading days, and $g(\cdot)$ denotes the clustering result transfer function.

Despite the effectiveness of the GMM in assessing the default risk of credit bonds, its results tend to be biased due to the non-normal distribution of default risks among bonds. Furthermore, this method produces non-smooth bond-specific default risk transitions over time, exhibiting erratic changes in certain intervals. Therefore, relying solely on the clustering method for labeling is impractical, and it is crucial to integrate other methods in order to ensure more stable, continuous, and reasonable sample labeling.

### *B. Market Index estimation*

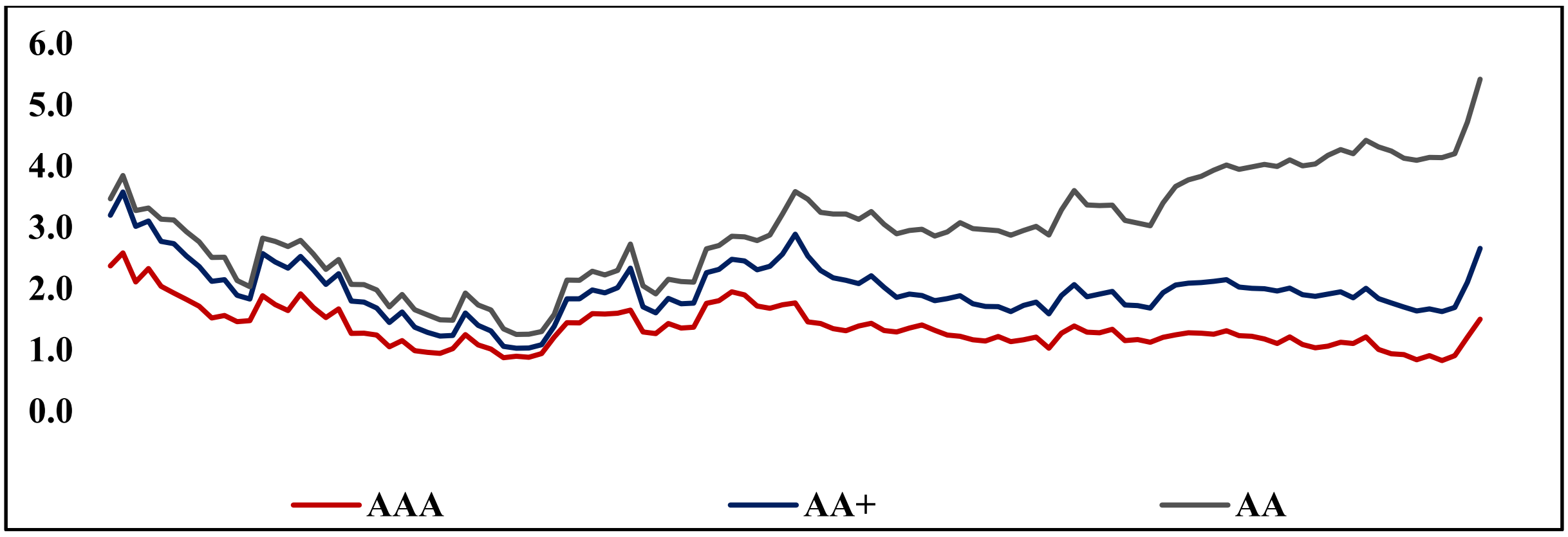

**Figure 5. Average credit spreads of different rated corporate bonds in China's credit bond market (AA-rated and above).**

Figure 5 illustrates that there is variation in the credit spreads of bonds with differing credit ratings, where lower credit spreads correspond to higher rated bonds and higher credit spreads correspond to lower rated bonds. In an equilibrium market, credit spreads reflect the expected probability of default (Hull, 2012). Therefore, the observed credit spreads can be used to estimate the unobserved default probability. The following equation is usually used to calculate the credit spread of bond $i$ at moment $t$:

$$cs_{it} = r_{it}^{b} - r_{t}^{f}, \tag{3}$$

where $r_{it}^{b}$ is the yield to maturity of bond at time $t$ and $r_{t}^{f}$ is the risk-free rate for the same period, which is represented by the treasury rate for the same period. At market equilibrium, the relationship between the bond default probability $p_{it}$ and the loss rate $D_{i}^{g}$ is as follows:

$$(1 - p_{it}) r_{it}^{b} - p_{it} D_{i}^{g} = r_{t}^{f}. \tag{4}$$

That is, the expected rate of return on the risky asset should be equal to the risk-free rate. From equations (3) and (4), it follows that:

$$p_{it} = \frac{r_{it}^{b} - r_{t}^{f}}{D_{i}^{g} + r_{it}^{b}} = \frac{cs_{it}}{D_{i}^{g} + r_{it}^{b}}. \tag{5}$$

The only variable lacking in equation (5) is the default loss rate $D_{i}^{g}$. Everbright Securities (2018) calculated the overall recovery rate of defaulted bonds to be 30.37% based on data from 2018. CICC (2022) counted and analyzed the weighted average recovery rate of 19.6% for all bonds that default by the end of 2022. Based on the study conducted by Chen et al. (2021), the recovery rate of bond defaults is uniformly set at 30% in this paper, and the default loss rate $D_{i}^{g}$ is all 70%. Since many credit bonds are not fully traded, there is often a lack of trading information that leads to high volatility in the yield to maturity and credit spreads calculated from trading prices. Additionally, we use a filtering technique to eliminate under- or over-estimates of default probabilities that may arise from the lack of trading information:

$$\hat{p}_{it}^{cs} = \begin{cases} 1, & \hat{p}_{it}^{\omega} \geq 1 \\ \hat{p}_{it}^{\omega}, & 0.05 \leq \hat{p}_{it}^{\omega} < 1 \\ 0.05, & \hat{p}_{it}^{\omega} < 0.05 \end{cases} \quad \text{and} \quad \hat{p}_{it}^{\omega} = \frac{cs_{it}^{\omega}}{r_{it}^{b} + 0.7} \tag{6}$$

where $cs_{it}^{w}$ is the value of credit spread $cs_{it}$ of bond $i$ at time $t$ smoothed by a moving average with window size $\omega$; $\hat{p}_{it}^{cs}$ represents estimated default probability $p_{it}$, which is derived using credit spread. This method provides a more accurate indication of the credit risk for bonds, however, it may still exhibit excessive volatility.

### *C. Default Trend Backward estimation*

Both of the above methods label default probabilities based on individual bond risk indexes. While they can objectively represent bond credit risk trends, their high volatility and instability are limiting factors that reduce the accuracy of day-to-day probability labeling. Filtering or moving average can help reduce excessive volatility but at the cost of lagging responses. Hence, we employ the Default Trend Backward method, leveraging information from defaulted and matured bonds in estimating credit risks while reasonably reflecting the variability of credit bond default probabilities.

**Default bond default risk trend backward estimation.** For bonds in default, the probability of default reaches 100% on the default date. Before the default date, issuers often attempt to stave off default using various means due to its substantial negative repercussions (Warner, 1977). This creates significant uncertainty regarding the bond's direction until the material default. A particular distribution governs the shift of bond default probability prior to the default date, thus allowing for the inversion of the probability trend if the default date is known. The trend derived from this method is an approximation of the collective traits shared by all defaulted bonds, unlike the above two methods which generate default probability trends specific to individual bonds.

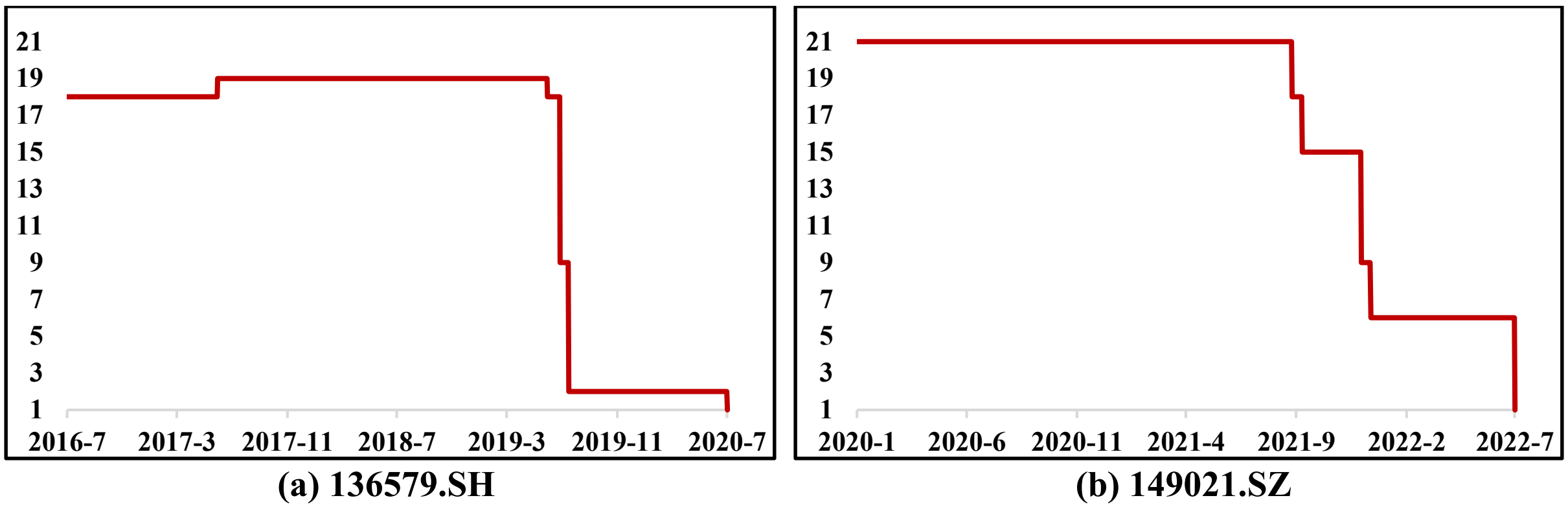

**Figure 6. Changes in credit ratings of defaulted bonds (two bonds as examples).**

Analyzing the evolution of the credit rating trend for defaulted bonds, we find that rating movements of most of these bonds are akin to the movements shown in Figure 6. More specifically, the credit ratings are relatively stable in the early stage of issuance, while in the later stage they show an accelerated decline until default occurs. Thus, we use this information to formulate an equation that describes the time-varying traits of default probabilities:

$$\hat{p}_{it}^{d} = \frac{N}{N + t_i}, \tag{7}$$

where $\hat{p}_{it}^{d}$ is the estimated backward default probability of default bond $i$ at time $t$, $t_i$ is the number of days until the default date; $N$ is the estimated number of days for default risk acceleration, which is set here to $120$ trading days (half of a year).

**Matured bond default risk trend backward estimation**. The default risk of matured bonds is also no longer uncertain as they do not default at or after maturity. Due to the minimal change in default probability for such bonds, overall trend in default probability is better explained by changes in credit rating. As a result, we use the linear interpolation for estimating the default probability of matured bonds based on credit rating changes:

$$\hat{p}_{it}^{m} = \frac{t_i(p_{iT} - p_{i0})}{T_i} + p_{i0}, \tag{8}$$

where $\hat{p}_{it}^{m}$ is the estimated backward default probability of matured bond $i$ at time $t$; $p_{i0}$, $p_{iT}$ are the initial default probability and maturity default probability of the bond, estimated from the issue rating and the latest rating before maturity, respectively; $T_i$, $t_i$ are the total maturity and remaining maturity of the bond.

## *D. Integrated Default Probability annotation*

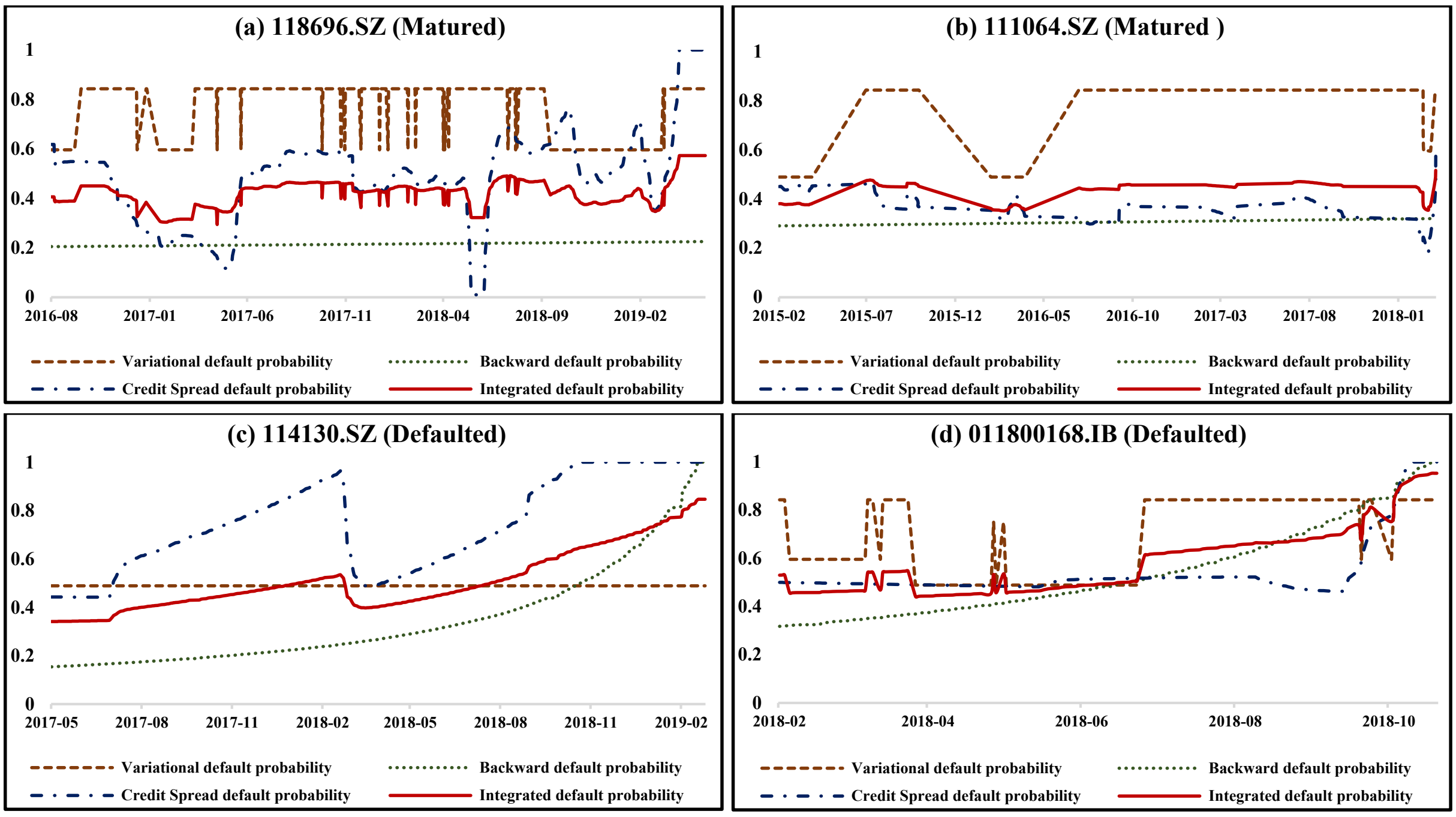

**Figure 7. Examples of daily frequency integrated default probability annotation results.**

We use the weighted average method to aggregate the bond default probabilities estimated by the above three methods into a comprehensive default probability. In turn, it can reflect both the individual characteristics and the overall trend of bond defaults, ensuring the reasonableness and accuracy of the annotation:

$$\hat{p}_{it} = \alpha^{gmm}\hat{p}_{it}^{gmm} + \alpha^{cs}\hat{p}_{it}^{cs} + \alpha^{b}\hat{p}_{it}^{b},\ b \in \{d,\ m\}. \tag{9}$$

The estimated integrated default probability $\hat{p}_{it}$ of the selected bond sample data is obtained from the variational default probability $\hat{p}_{it}^{gmm}$, the credit spread default probability $\hat{p}_{it}^{cs}$ and the backward default probability $\hat{p}_{it}^{b}$, $b \in \{d,\ m\}$ by weighting the average of the weights $\alpha^{gmm}$, $\alpha^{cs}$, $\alpha^{b}$. $\hat{p}_{i(t-1)}$ is used as the prior default probability of the bond at the time $t$, and the prior default probability of the first period is initialized to 0.5.The weights are set to 0.3, 0.3 and 0.4, respectively.

Figure 7 illustrates the results of the composite default probability markers for some bonds. It can be seen that the weighted default probabilities estimated by the three methods are more stable, which is a greater annotation under the existing data conditions. We establish a database of Chinese credit bonds with daily frequency default probability labels, and lay the foundation for the training and prediction of deep neural networks.

## IV. Construction and Training of Default Probability Prediction Model

The correlation of bond data in the time series dimension is obvious. Indexes belonging to the same dimension in the index system also tend to have a large correlation, and they jointly reveal a major aspect of factors affecting bond defaults. To fully extract the time-series information and cross-sectional indicator correlation from bond data, we construct a default probability prediction model based on Convolutional Long and Short-term Memory (ConvLSTM) deep neural network. This section first introduces the structure and principles of ConvLSTM, then shows the structure of the proposed default probability prediction model in detail, and finally illustrates the model training method.

### *A. ConvLSTM*

Hochreiter and Schmidhuber (1997) proposed the LSTM based on RNN to selectively reinforce important information and forget unimportant information by adding a nonlinear gating structure inside the network neurons, thus effectively avoiding the gradient disappearance problem during the training of long sequences and ensuring the effectiveness of temporal information transfer. However, the main drawback of LSTM is that it cannot fully extract the spatial features of the data by the fully connected operation in the coding process. To address this problem, Shi et al. (2015) proposed the ConvLSTM model, which effectively mines the 2D spatial information of images by introducing convolutional operations in the encoding process, while ensuring the reliability of temporal information with the help of a gating mechanism:

$$\begin{aligned}
i_t &= \sigma(Conv(W_{xi}, X_t) + Conv(W_{hi}, H_{t-1}) + W_{ci}C_{t-1} + b_i) \\
f_t &= \sigma(Conv(W_{xf}, X_t) + Conv(W_{hf}, H_{t-1}) + W_{cf}C_{t-1} + b_f) \\
C_t &= f_t C_{t-1} + i_t \tanh(W_{xc}X_t + W_{hc}H_{t-1} + b_c) \\
o_t &= \sigma(Conv(W_{xo}, X_t) + Conv(W_{ho}, H_{t-1}) + W_{co}C_t + b_o) \\
H_t &= o_t \tanh(C_t)
\end{aligned} \quad . \tag{10}$$

$Conv(\cdot)$ denotes the convolution operation; $i_t$, $f_t$, $C_t$, $o_t$, $H_t$ represent the input gate, forgetting gate, memory module, output gate and hidden state, respectively. ConvLSTM differs from the traditional LSTM in its operation of weight parameters and feature vectors. It employs convolution instead of dot product, while maintaining similar temporal operation steps. Such change helps address the traditional LSTM's weakness in handling spatial information. This development can allow for the construction of neural networks to be tailored for the specific modeling requirements of financial data.

## *B. Chinese Credit Bond Default Probability Prediction Model Architecture*

**Model Architecture.** Based on the comprehensive analysis, we develop a prediction model to estimate the default probability of Chinese credit bonds, as illustrated in Figure 8. The input data is daily frequency feature time series $[X_1, X_2, \cdots, X_w]$ of individual bonds in the previous period with lag window size $w$, where $X_t = [x_{t,1}, x_{t,2}, \cdots, x_{t,53}]$ denotes the features of individual bond samples at the time $t$. The sequence convolution is firstly conducted to extract the correlation of each index. Then a flatten operation is executed, followed by the acquisition of the features embedded with integrated temporal and spatial information through 10-layer stacked LSTM processing. Finally, the fully connected layer is utilized to obtain the probability of predicting default risk.

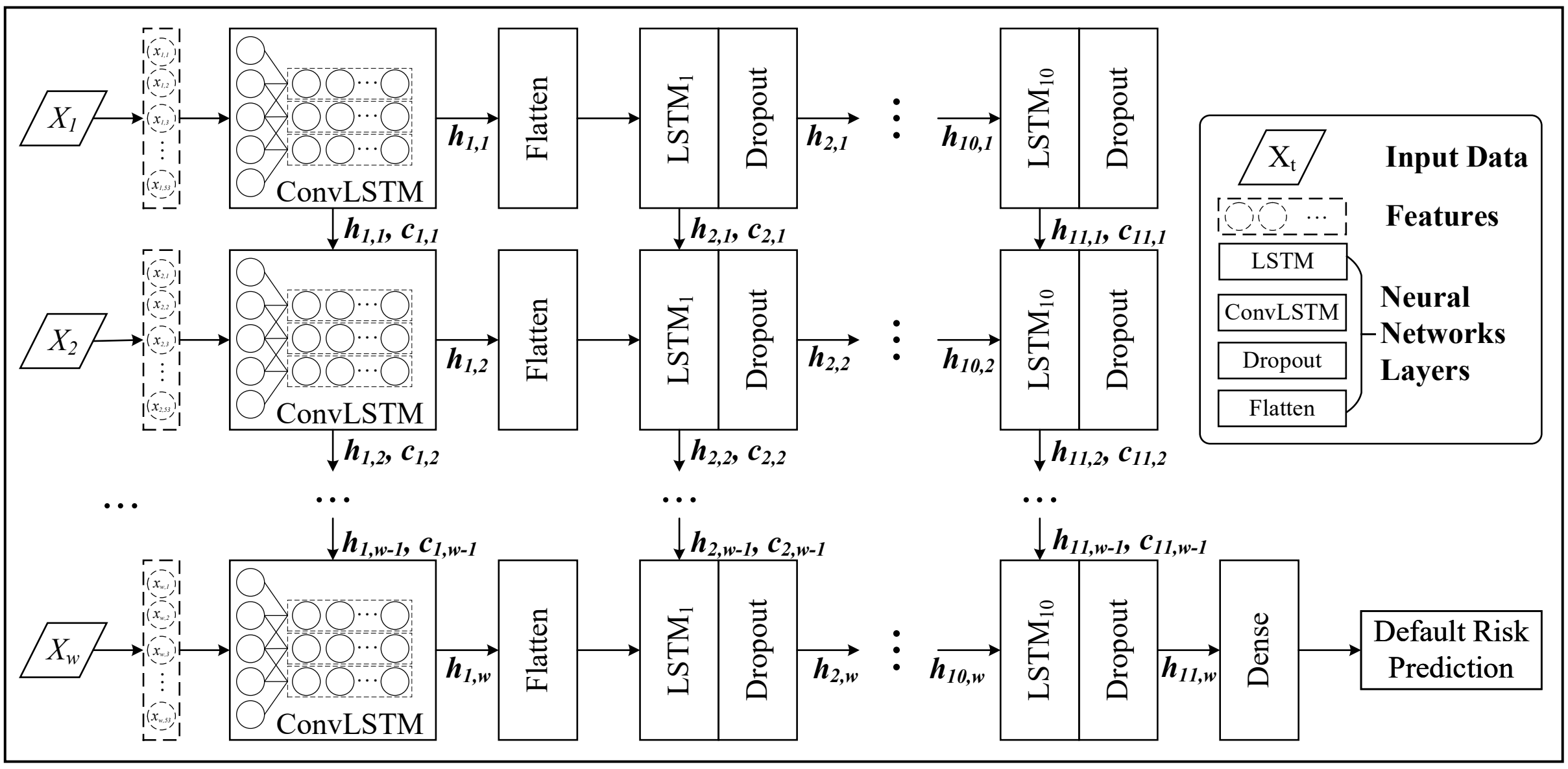

**Figure 8. The architecture of our Chinese Credit Bond Default Probability Prediction Model.**

We introduce ConvLSTM in the first layer of encoding, and subsequently, used the traditional LSTM for data processing instead of adopting this structure in every layer, as demonstrated by Shi et al. (2015). This is because the number of features belonging to the same dimension is limited, and numerous convolution operations can cause irrelevant features to interfere with each other. Moreover, it aims to reduce the complexity of model operations, and consequently decrease the training and inference time. The purpose of introducing the Dropout layer is to prevent every bit of information from each part to pass to the next layer and to minimize the chances of overfitting. The dropout rate of the Dropout layer after the first three LSTM layers is 0.5, the last layer is 0.125, and the rest are 0.25. Considering the feature obtained by the latter neural network layer is a high-dimensional generalization of the information in the bottom layer, the dropout rate should be decreased.

**Loss Function.** The ultimate goal of this model is to predict the next-day default probability by leveraging historical sequence information of individual bonds with window size $w$. Therefore, accuracy is an important optimization objective. In this paper, we adopt the commonly used Mean Square Error (MSE) in current time series forecasting as the loss function, following the approach of Jin et al. (2019):

$$L = \frac{1}{N}\Sigma_{i=1}^{N}\frac{1}{T_i}\Sigma_{t=1}^{T_i}\left(\hat{p}'_{it} - \hat{p}_{it}\right)^2, \tag{11}$$

where $N$ is the total number of credit bonds, $T_i$ is the total number of days to predict the default probability for credit bond $i$, and $\hat{p}'_{it}$ is the default probability prediction value. The RMSProp algorithm proposed by Cho et al. (2014) is utilized to optimize the model parameters. Default values of relevant keras modules serve as the hyperparameters used.

## *C. Model Training*

**Dataset Design and Preprocessing.** In the Chinese credit bond database constructed above, which contains daily default probability labels, some of the credit bond indexes exhibit incompleteness and inconsistent magnitudes. Therefore, we preprocess the data by two ways. 1) Using **linear interpolation** to fill in missing values. The approach is preferred over removing missing values, which would have disrupted the continuity of time series data. 2) Different indexes have large differences in statistical units and inconsistent fluctuations, which can adversely destabilize the model, so the **mean-variance standardization** method is used to standardize individual indicators for each individual bonds:

$$x^*_{t,f} = \frac{x_{t,f} - \mu_f}{\sigma_f}$$

$$\mu_f = \frac{1}{T}\Sigma_{t=1}^{T}x_{t,f}$$

$$\sigma_f = \sqrt{\frac{1}{T}\Sigma_{t=1}^{T}(x_{t,f} - \mu_f)^2}, \tag{12}$$

where $x_{t,f}$ is the original value of the indicator $f$ at time $t$; $x^*_{t,f}$ is the standardized value; $\mu_f$ and $\sigma_f$ are the mean and variance of the index in the total period. This eliminates the differences caused by different units and greatly reduces the magnitude of data fluctuations.

The preprocessed data contains both time-series and spatial information. We first construct input features of dimension $(N_i, w, 53)$ for each individual bond using the window sliding method, and labels of corresponding dimension $(N_i, w, 1)$. The lag window size $w$ is a hyperparameter that will be discussed in detail later. Then the high-risk (default and below A-rated) and low-risk (matured) credit bonds are divided into training, validation, and test sets in an 8:1:1 ratio. Finally, the SMOTE method proposed by Chawla (2002) is used to balance the sample size of various types of credit bonds, with a focus on resampling high-risk bonds in the training set to avoid model training bias from a too large proportion of matured credit bonds. And the original information is retained to the maximum extent without changing the data distribution.

**Training Environment and Evaluation Metrics**. The proposed model for predicting default probability and its comparative analysis are developed and trained using the Python 3.8-based Keras framework, with TensorFlow 1.15.5 and Cuda 11.4. The training involved a Batchsize of 2 using a Windows operating system with an RTX3080Ti GPU. In this study, we employ the **Root Mean Square Error** (RMSE) and **Mean Absolute Error** (MAE), which are usually used in the time series forecasting field, as evaluation criteria to determine forecast accuracy. The underlying principle is to compute the difference between the predicted probability of default and the actual value; smaller values for either RMSE or MAE indicate higher accuracy of prediction.

## V. Prediction Results and Evaluation

### *A. Comparative Analysis of Model Test Results*

In this paper, we compare the accuracy of our proposed model for default probability prediction with four other methods. **(1) Boosting** model, which has the characteristics of low bias and stability, and is one of the most widely used machine learning methods in various fields. To compare the performance of our proposed model with the traditional temporal neural network architecture, we construct **(2) RNN**, which is the basic temporal neural network architecture and is representative in the processing of time series. **(3) LSTM** model, a deep neural network consisting of 10 stacked LSTM layers following the same dropout layer, which should have stronger recognition ability on temporal features. **(4) Pure ConvLSTM** (PConvLSTM) model is also built, where all temporal information processing modules are LSTM layers utilizing convolutional operations.

To ensure the validity of our model test results, we acquire the experimental data exclusively from the test set. As such, SMOTE is not apply to the test set data, which upholds the realistic distribution of the data in real-world scenario. Due to the difficulty in determining the size of the lag window directly, considering the realistic trading scenarios, we select four cases $w=2$, $w=5$, $w=7$, and $w=10$ for testing. The experimental results are presented in Table II. It can be seen that regardless of the window size, our method has the smallest RMSE and MAE. Both the Boosting and LSTM demonstrate better prediction accuracy for all window sizes. Conversely, the PConvLSTM model exhibits high accuracy in the small window size range, which steadily decreases with increasing lags.

**Table II**

**Experimental Results for Next-Day Default Probability Prediction**
**(The top-2results are highlighted in bold and underlined)**

| Method | $w=2$ | | $w=5$ | | $w=7$ | | $w=10$ | |
|---|---|---|---|---|---|---|---|---|
| | RMSE | MAE | RMSE | MAE | RMSE | MAE | RMSE | MAE |
| Boosting | 0.066 | 0.049 | 0.066 | 0.046 | 0.065 | 0.047 | 0.065 | 0.048 |
| RNN | 0.154 | 0.137 | 0.150 | 0.123 | 0.247 | 0.188 | 0.156 | 0.147 |
| LSTM | 0.071 | 0.050 | 0.069 | 0.049 | 0.070 | 0.050 | 0.063 | 0.046 |
| PConvLSTM | 0.064 | 0.048 | 0.065 | 0.050 | 0.078 | 0.060 | 0.081 | 0.079 |
| Ours | **0.058** | **0.041** | **0.063** | **0.045** | **0.062** | **0.043** | **0.063** | **0.046** |

The experimental results indicate the following: (1) The intelligent prediction model of Chinese credit bond default risk based on ConvLSTM incorporates both cross-sectional and time-series information, and outperforms the Boosting model that emphasizes cross-sectional information, as well as the RNN and LSTM models that focus on time-series information. (2) Our model is more stable than PConvLSTM, which exhibits significantly lower prediction accuracy at longer lags. This suggests that the deep convolution operation tends to introduce irrelevant variables that interfere with each other. This interference decreases the model accuracy as the lag period increases and the data features become more complex. (3) The model performs optimally when the window size is 2, which indicates that the two-period lag has been able to characterize the time-series correlation to a greater extent in this dataset. Of course, the selection of the lag window size is related to both the amount of data and the model's complexity, and a matching configuration is necessary for high model prediction accuracy.

## *B. Evaluation of day-by-day Default Probability Forecast Results*

The China bond market implied rating - Bond rating, which is discovered by China Bond Financial Valuation Center from the financial information of issuing entities and market movements, can dynamically reflect the credit risk of bonds. It is currently the most extensive and timely adjusted rating in China's credit bond market. In this paper, the obtained forecast results are compared with is. The discrete rating data is transformed into continuous default probabilities by applying equation (3). Then, the default probabilities are compared to the results predicted by our model. This validation confirms the usefulness and forward-looking nature of this model for estimating day-by-day default probabilities.

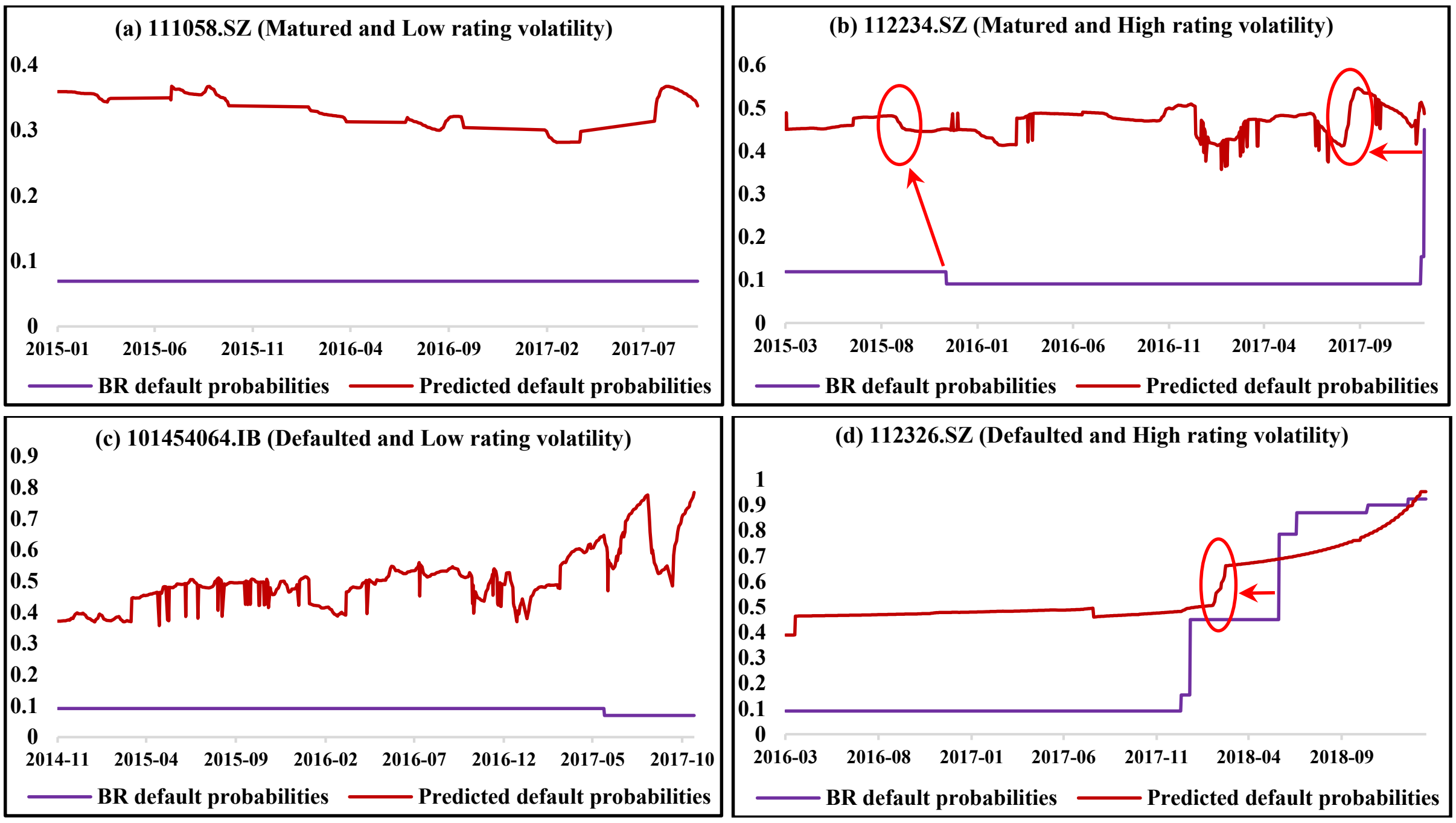

**Figure 9. Example of predicted default probability versus default probability converted from Chinese bond ratings ($w = 2$).** The **red arrows** depict that our model provides an early warning for risk changes.

Comparing the predicted result of all samples in test set with the default probabilities obtained from the bond ratings (BR default probabilities), a linear regression yields an $\boldsymbol{R^2}$ **of 0.62**. This indicates a significant correlation between the Chinese bond ratings and the predicted daily default probabilities at an aggregate level, which provides a reasonable picture of the credit risk of bonds. Nonetheless, the Chinese bond ratings are persistently inflated and lagging, which not only leads to an underestimation of default probabilities but also results in an untimely measurement of risk. Our approach improves this significantly.

Figure 9 provides a comparison between the predicted results with a window size of 2 and the Chinese bond ratings, including representative examples of maturities and defaults with less and more volatile ratings. The "BR default probabilities" is the continuity probability transformed by the function, whereas the "Predicted default probabilities" depicts the result of daily default risk forecast of individual bonds by our model. The figure reveals the following observations: (1) The prediction results corroborate the overall trend of credit risk reflected by the Chinese bond ratings, which fortifies the soundness of our framework. (2) The predicted default probability is relatively higher, implying that our model has mitigated the inaccuracies that stem from the Chinese bond ratings, and as such, the prediction results better align with the actual risk level. (3) Our model exhibits a robust early warning capacity for sudden changes in default risk. For instance, for the two bonds illustrated in Figure 9.(b) and 9.(d), there is a predicted default probability shift, about 2 to 3 months earlier than the changes in the bond ratings.

## VI. Conclusions and Insights

The material default of "11 Chaori Bond" marked the end of the "rigid payment" situation that persisted in the credit bond market. Since then instances of credit bond defaults have gradually increased and credit risks have become increasingly prominent. A thorough analysis of credit bond defaults indicates that sudden defaults can be attributed to several factors which may include external macro and industry shocks, as well as internal weaknesses in investment, management and financing. In a credit bond market where information disclosure is incomplete, investors often find it challenging to evaluate bond credit risk effectively. To this end, we propose an intelligent forecasting framework for default risk to fully explore information based on accessible structured data. In turn, the default probability is reasonably predicted, thereby enabling investors to anticipate risks in a timely manner and foster the sound and stable growth of China's credit bond market.

The credit bond risk index system that we developed effectively characterizes individual bond credit and has proved to be useful in analyzing many default cases. The comprehensive annotation of daily frequency default probability can reflect both the overall default risk characteristics of the bond market and the variation characteristics of individual bond default risk. It provides a precise insight into the historical default probability of matured and defaulted bonds. We conduct extensive experimental analyses that demonstrate that our model has a significantly higher prediction accuracy than other frequently used methods. After comparing with the default probability reflected by the bond rating, on the one hand, we find that our model can effectively avoid the strange appearance of inflated ratings; on the other hand, the model is forward-looking and can warn changes in credit risk in a more timely manner for both matured and defaulted bonds.

Based on the conclusions above, this paper gets the following insights on the regulation of credit bond market and the warning of default risk. Firstly, regulators should enhance the information disclosure mechanism, bond issuance guidelines and binding provisions due to the interest-based relationship between rating agencies and issuers, the significant information asymmetry between issuers and investors, and the complexity of credit risk generation and characterization. Regulators must guarantee an effective environment that enables efficient default risk identification. Secondly, rating agencies ought to further improve the independence and objectivity of their ratings, while also improving the timeliness and accuracy of their ratings, so as to effectively inform investors of the risks, thereby avoiding large shocks caused by sudden defaults of highly rated bond. Thirdly, issuers should continue to improve their corporate governance capabilities, concentrate on operations, investment and financing, and enhance the ability to withstand external shocks. Fourthly, investors should scrutinize information throughout the process of risk accumulation and evolution to establish an effective analysis system for timely and effective risk monitoring. Finally, given the rising exposure to credit risks, China's credit bond market should establish a diversified default risk-sharing mechanism and refinancing mechanism to mitigate negative impacts of defaulted bonds on the market and investors. This will also prevent concentrated defaults from leading to systemic risks.